\documentclass{article}

\usepackage[english]{babel}

\usepackage[letterpaper,top=2cm,bottom=2cm,left=1.5cm,right=1.5cm,marginparwidth=1.75cm]{geometry}
\usepackage{natbib}
\usepackage{amsmath}
\usepackage{graphicx}
\usepackage[colorlinks=true, allcolors=blue]{hyperref}
\usepackage{caption}
\usepackage{subcaption}
\usepackage{setspace}
\usepackage{authblk}
\usepackage{blindtext}
\usepackage{multicol}
\usepackage{authblk}

\usepackage{xcolor}

\title{Role of the ocean for fast atmospheric evolution revealed by machine learning}

\author[a]{Bobby Antonio\thanks{bobby.antonio@physics.ox.ac.uk}}
\author[a,b]{Kristian Strommen}
\author[a]{Hannah M. Christensen}

\affil[a]{Atmospheric, Oceanic and Planetary Physics, University of Oxford, Sherrington Road, Oxford, OX1 3PU, United Kingdom}
\affil[b]{European Centre for Medium-Range Weather Forecasts, Shinfield Rd, Reading, RG2 9AX, United Kingdom}

\begin{document}

\maketitle

\begin{abstract}
There have recently been many efforts to create machine learnt atmospheric emulators designed to replace physical models. So far these have mainly focused on medium-range weather forecasting, where these ‘Machine Learnt Weather Prediction' (MLWP) models can outperform leading operational forecasting centres. However, because of this focus on shorter timescales, many of these emulators ignore the effects of the ocean, and take no ocean variables as inputs. We hypothesise that such MLWP models have learnt a best-guess of the evolution of the atmosphere, by implicitly inferring ocean conditions from atmospheric states, with no access to ocean data. Turning this limitation into a strength, we use it as a means to study the role of the oceans on the evolution of the atmosphere. By exploring how model forecast errors relate to properties of the air-sea interface, we infer what ocean information these atmospheric emulators are able to derive from atmospheric data alone, and what they cannot. This highlights the regions and processes through which the ocean independently influences the atmosphere on fast timescales. We perform this analysis for GraphCast, finding clear relationships between air-sea properties and the forecast errors over the ocean, including clear seasonal effects. We then explore what this reveals about GraphCast's internal representation of the ocean. In addition to understanding real-world ocean-atmosphere interactions, this analysis provides guidance for improving forecast skill and physical realism in MLWP models, and for informing how future machine learning models should use ocean information on short timescales.
\end{abstract}

The processes that exchange heat, mass and momentum between the ocean and atmosphere drive changes in atmospheric circulation, cloud cover, and ocean currents \citep{chelton_coupled_2010}. Studying and understanding the nature and timescales of these ocean-atmosphere interactions is therefore crucial in improving our understanding of the Earth system and in developing models that can produce skilful long-range forecasts and faithfully represent climate variability.

Much effort has therefore been applied to characterising ocean-atmosphere interactions. A typical method to do so is via the use of simple stochastic coupled differential equations, based on the models of Hasselmann and Frankignoul \citep{hasselmann_stochastic_1976, frankignoul_stochastic_1977}, and developed further by \cite{barsugli_basic_1998} and \cite{wu_local_2006}. Through the analysis of lagged correlations between sea surface temperatures (SSTs) and surface fluxes or surface wind stress, the driver of the coupling variability can be inferred \citep{storch_signatures_2000}. Whilst this gives an intuitive method to investigate the air-sea interaction, its limitation is that it requires making simplifying assumptions and approximations, including a linearisation of the heat flux and long-wave radiation terms. In applications, it also typically relies on the availability of heat flux data, which is either sparse or relies on bulk formulae to calculate from observations and reanalysis.

An alternative method to probe the air-sea interaction is to perform experiments with atmospheric general circulation models (AGCM) with different limits placed on the interaction with the ocean. By comparing the performance of such models when forced with persisted anomalies, observed SSTs or coupled to an ocean model, it is possible to infer the effects of the ocean interaction on the atmosphere. For Numerical Weather Prediction (NWP) models, the analysis in \cite{mogensen_effects_2018} showed how coupling to an ocean model for short-range forecasts could improve 2-metre temperature forecasts in a specific event in the Gulf of Lions, whilst \cite{takaya_implementation_2010} investigated the effects of coupling the IFS to a simplified ocean model. \cite{mogensen_tropical_2017} investigated how coupling NWP forecasts to an ocean model can help improve errors in forecasting the intensity of tropical cyclones, relative to runs using persisted SST anomalies. Subsequent work in \cite{vellinga_evaluating_2020} explored spatial differences in errors for a coupled and uncoupled NWP system out to 168hrs lead time, as well as performing a more extensive analysis of the effect of air-sea coupling on tropical cyclones. Similar approaches have also been used for model runs on longer timescales \citep{kushnir_atmospheric_2002, wu_local_2006}. Such a setup is reliant on a well-initialised ocean state, and accurately capturing the behaviour of ocean dynamics and air-sea interactions, which can be complex due to the turbulence in the boundary layer.

Recent years have seen a surge in successful applications of machine learning to tackle problems in weather and climate. Of particular note is the development of Machine Learnt Weather Prediction (MLWP) models that rival the skill of conventional NWP models on certain metrics. So far, these models have focused mainly on short- to medium-range weather forecasts, which means the models can achieve state-of-the-art skill without incorporating any information about the slow varying components of the Earth-system including the ocean. One example of such a model is GraphCast \citep{lam_learning_2023}, which receives no information about the ocean, land, or cryosphere, but at the time of its publication achieved state-of-the-art forecast skill. 

If we denote $X$ as the atmospheric state vector, and $Y$ as the state vector that describes other components of the Earth system (e.g.~ocean and cryosphere), then MLWP models such as GraphCast that are trained solely on atmospheric fields can, at best, learn an approximation of the dynamics $x(t) = f(x)$, where the lowercase $x$ indicates a subset of the true state vector $X$ (e.g.~at lower resolution, and on a number of discrete vertical levels). We hypothesise that the function $f(x)$ is an estimate of 
$\mathcal{F}(x) = \int F(x,Y) p(Y |x) dY$ where $F(X,Y)$ represents the true coupled dynamics; this is the most likely evolution of the atmosphere given knowledge only of the atmosphere, integrating over the possible effects of the other unknown Earth system components. Because the missing variables, $Y$, typically vary slower than the atmospheric $x$ variables, it is hoped that they will not change significantly during the MLWP forecast such that a skilful prediction may still be made. Nevertheless, any MLWP prediction will contain an error due to the incomplete knowledge of the atmospheric state, errors in the estimate, $f(x)$ of $\mathcal{F}(x)$, and the lack of knowledge about the state vector $Y$. Here we focus on the latter source of error and hypothesise that, in learning skillful predictions of the weather without sea surface information, GraphCast must make an uninformed guess about the ocean dynamics, and therefore will have greater forecast error in areas where the unseen ocean component drives the dynamics.

We investigate this hypothesis by exploring the relationships between GraphCast's errors and several variables relevant to the sea surface, demonstrating that there is a recognisable signal of the ocean in these errors consistent with previous analysis of ocean-driven atmospheric dynamics. A clear seasonal behaviour can be observed, with stronger ocean-driven dynamics observed in the northern Pacific and northern Atlantic over the boreal summer compared to the winter, and a clear signature of the lagged response of the ocean to heating during the summer. The locations of the areas of the ocean that have the highest impact on GraphCast's errors align with previous results for longer timescales using coupled linear models, but with differences in the strength of interaction indicated by the two methods. This idea is then explored further by training a linear regression model to predict the ocean-relevant component of GraphCast's errors, to assess the relative magnitude of these errors relative to other model errors. For 24-h lead time forecasts, we find that the linear model leads to a 1\% reduction in the error, which to our knowledge provides the first quantitative estimate of the impact of SST on forecast skill at these short timescales. 

Relative to the lagged correlation and AGCM approaches, we see there being two main benefits in our approach. Firstly, GraphCast is trained without any flux data, so we are able to identify regions where the SST heating the atmosphere is important using sea surface temperature and 2-metre temperature data only. Therefore we can potentially leverage a simpler set of observational data other than flux measurements to infer the air-sea interactions. Secondly, this approach is not reliant on constructing a simplified model of the dynamics, which may be too simple and can be cumbersome to interpret (a similar argument applies to the approach in \cite{yang_causal_2024}, which also assumes a linear model). Thirdly, we are not reliant on a well-initialised ocean state, or accurately capturing the physical equations of air-sea interactions, which can be complex due to the turbulence in the boundary layer. 

 Our approach therefore provides a novel and potentially more accurate way to identify ocean-driven air-sea interactions at short timescales, to highlight processes that are particularly important for accurate short-range forecasts, and to shed light on where the benefits of a coupled forecasting system may come from. An understanding of how GraphCast represents ocean dynamics could also aid in improving the future training of models, particularly as machine learning models begin to be rolled out to longer forecast horizons, such as seasonal \citep{kent_skilful_2025, zhang_advancing_2025, antonio_seasonal_2025} or climatological \citep{kochkov_neural_2024, cresswell-clay_deep_2024, watt-meyer_ace2_2025, clark_ace2-som_2025} timescales, where a realistic representation of the air-sea interaction becomes crucial.

\section*{Results}
\subsection*{Impact of ocean on 2-metre temperature errors}
\label{sec:graphcast_2mt}

In this section, we look at how the errors in GraphCast's T2m forecasts relate to properties of the air-sea interface. We choose T2m as the target variable since it is strongly influenced by the ocean, and it is weighed most heavily in GraphCast's loss function during training. We compare T2m with both SST and SST-T2m, as the effect of the ocean on the atmosphere is approximately a function of SST-T2m, SST and $\text{SST}^4$ (see Materials and Methods). To compare targets and inputs we use Spearman correlation, which correlates the rankings of the input and target variables rather than the absolute values; this permits nonlinear relationships to be captured more accurately compared to Pearson correlation. Note however that this introduces subtleties when comparing to previous works such as \cite{sun_seasonality_2021} and \cite{bishop_Scale_2017}, where covariances are used, since the correlations will be normalised relative to local variability whereas the covariances are not.

We consider GraphCast errors at a 24-hour lead time, in order to avoid diurnal effects. The 2-metre temperature error $\varepsilon_T$ is defined as
\begin{align}
    \varepsilon_T = T_{gc} - T_{ERA5}
\end{align}
where $T_{gc}$ is GraphCast's 2-metre temperature forecast, and $T_{ERA5}$ is the 2-metre temperature from ERA5. These errors are then compared to daily-averaged predictors from ERA5 (see Materials and Methods).

\begin{figure*}[!ht]
    \centering
        \includegraphics[width=17.8cm]{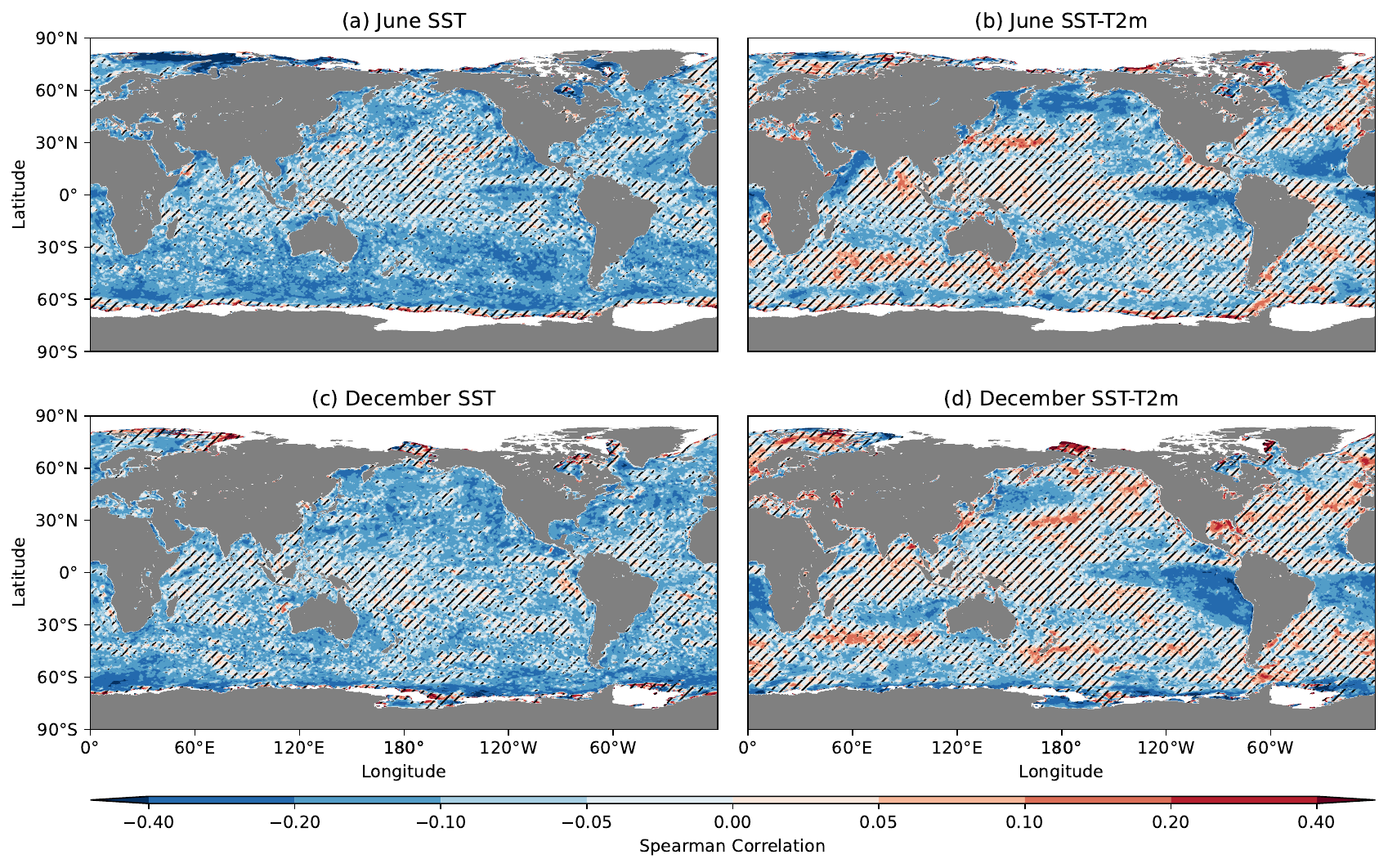}
       \caption{Spearman correlation between GraphCast's T2m error at 24hrs lead time and variables at the sea surface. Hatching indicates where the correlation is not significant according to a two-tailed t-test at the 95\% confidence level. (a) SST for June, (b) SST-T2m for June, (c) SST for December, (d) SST-T2m for December. Note that the colorbar increases exponentially.}
    \label{fig:t2merr_cov}
\end{figure*}

We start by investigating the relationships between $\varepsilon_T$, SST, and SST-T2m. The Spearman correlations between these variables are shown for June and December 2004-2013 in Fig.~\ref{fig:t2merr_cov}, with hatching indicating areas where the correlation is not significant (see Materials and Methods). For both June and December, SST shows significant negative correlation with $\varepsilon_T$ over large regions of the ocean, whilst SST-T2m shows a smaller area with significant correlations, but with areas of more concentrated, higher correlation. In June there is particularly strong negative covariance between $\varepsilon_T$ and SST-T2m over the northern Pacific, tropical eastern Pacific, in the Labrador Sea, off the western coast of Africa, and in the Arabian Sea. These ocean-driven dynamics in the Arabian sea were also noted in e.g.~\cite{sun_seasonality_2021} using lagged correlations, which they attribute to monsoon-driven upwelling. In December there are particularly noticeable correlations with SST-T2m around the Kuroshio Extension, tropical eastern Pacific, and southern Atlantic ocean. It is notable that, whilst in June there are high correlations for both SST and SST-T2m in the eastern Pacific, in December high correlations are only seen for SST-T2m. The areas of high correlation also bear a resemblance to the areas of improved skill seen for a coupled NWP forecast relative to an uncoupled one in Fig. 2(b) of \citep[e.g.][Fig 2(b)]{vellinga_evaluating_2020} (where some differences are to be expected since the correlation is normalised relative to the local variability).

The predominantly negative sign of the correlations in Fig.\ref{fig:t2merr_cov} (a)-(d) is intuitive; if a region has a positive SST anomaly, the real atmosphere should receive more heat than usual. However, since GraphCast does not  have information about this anomaly, the forecasts produced by the model will have a cool bias, hence a negative correlation. Exceptions to this can be seen particularly in the southern edge of the Kuroshio Extension in the boreal summer (Fig.\ref{fig:t2merr_cov} b), and around the Agulhas Return Current in the austral summer (Fig.\ref{fig:t2merr_cov} d); these suggest areas where a high SST-T2m driven by ocean variability results in lower surface temperatures, perhaps by feedback mechanisms, such as cloud formation or advection, that result in an overall cooling. These effects also appear to be present, albeit to a much lesser extent, during the respective winters of each of these currents (i.e.~panel (d) for the Kuroshio Extension and panel (b) for the Agulhas Return Current). 

Many of the areas of larger correlation seen for SST-T2m coincide with areas of high oceanic vertical eddy heat transport diagnosed in \cite{jing_maintenance_2020} using the Community Earth System Model. Note that it is hard to directly compare the correlation plots in Fig.~\ref{fig:t2merr_cov} with plots of vertical eddy transport, since correlation is normalised with respect to the local variability, whereas the eddy transport values are not. Nonetheless, the high values of vertical eddy transport in the Arabian sea in boreal summer, around the Kuroshio Extension in boreal winter, and in the eastern Pacific are all reflected in areas of high correlation. In the eastern Pacific in boreal winter, the vertical eddy transport increases relative to the boreal summer, which also appears to be reflected in the increased correlations in the eastern Pacific from Fig.~\ref{fig:t2merr_cov} (a) to (d). A notable difference can be seen around the Gulf Stream in boreal winter, for which \cite{jing_maintenance_2020} show an increase in vertical eddy heat transport, whereas Fig.~\ref{fig:t2merr_cov} (d) shows a decrease in correlation for December.

For June [Fig.\ref{fig:t2merr_cov}(a)] we observe a stronger relationship with SSTs in the southern hemisphere, which we interpret being due to the lagged response of the ocean to heating during the boreal summer, such that heat is released from the ocean into the atmosphere in the boreal winter. The reverse is seen in December, where there are more significant correlations between $\varepsilon_T$ and SST in the Northern hemisphere.

\subsection*{Comparison to lagged correlation method}

We make a direct comparison to the technique of lagged correlations by calculating lagged correlations between sensible heat flux anomalies and SST anomalies (both taken from ERA5). Several works have detailed, under some assumptions, how these lagged correlations can be interpreted. In order to analyse these lagged correlations, we follow the characterisation of them in \cite{bishop_Scale_2017}, as solutions to a simplified linear equation of coupling between atmosphere and ocean (see Materials and Methods); where the ocean drives the atmosphere, the lagged correlations between SST anomalies and the surface heat flux will have high magnitude that does not change sign with the sign of the lag. Other works also determine the ocean-driven dynamics based on the behaviour of the SST tendency \citep{bishop_Scale_2017,wu_local_2006,wallace_spatial_1990}. However, our analysis based on SST tendencies at this timescale revealed very few regions where the sign of the SST tendency correlation was significant and changed sign between positive and negative lags. 

Note that, unlike \cite{bishop_Scale_2017}, we follow ERA5's convention of positive heat fluxes going downwards, therefore typical correlations between SST and heat fluxes when the ocean drives the dynamics are expected to be negative (i.e. high SST anomalies result in more heat flux going upwards, corresponding to more negative heat flux anomalies). This allows an easier visual comparison with Fig.~\ref{fig:t2merr_cov}.

\begin{figure*}[!ht]
    \centering
        \includegraphics[width=17.8cm]{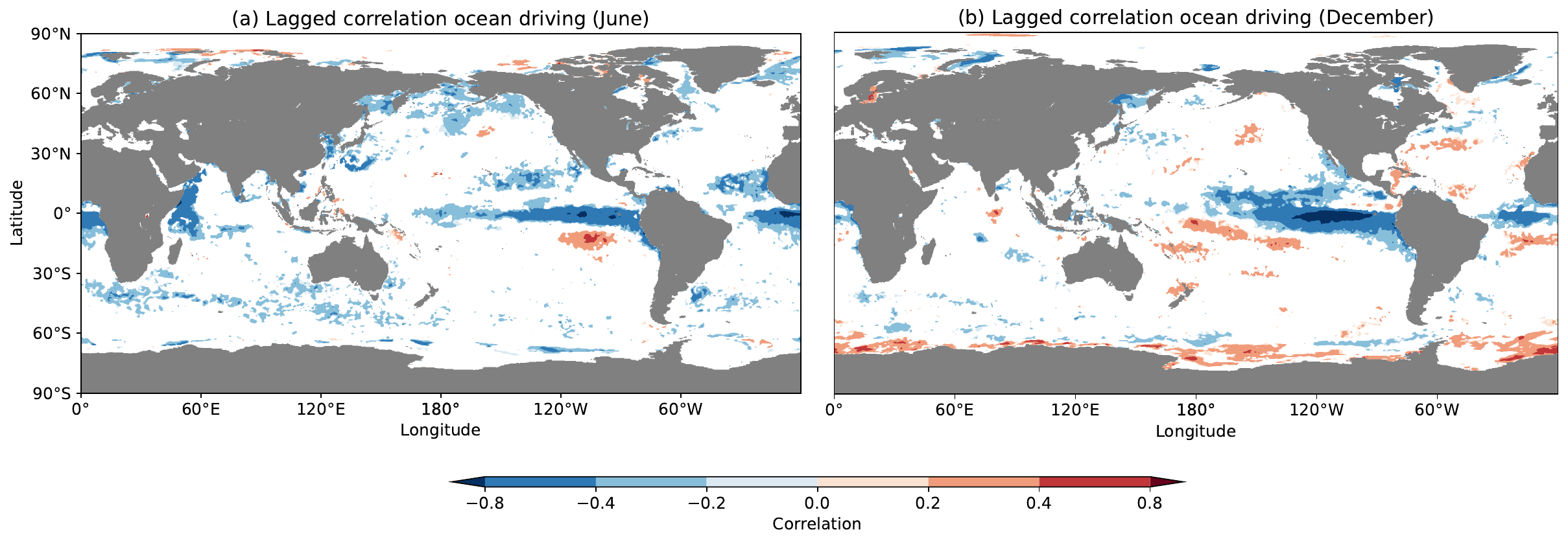}
       \caption{Inferred areas of ocean-driven dynamics, using the technique of lagged correlations between the observed SST anomaly and mean surface heat flux anomaly. Red (blue) colouration indicates where there is significant positive (negative) correlation that does not change sign with lagging, where significance is assessed according to a two-tailed t-test at the 95\% confidence level. (a) over June months and (b)  over December months. Correlation is shown only for the points that have significant correlation at zero lag, and where the correlation does not change sign between a lead/lag of 72 hours. Note the colorbar increases exponentially.}
    \label{fig:sst_anom_vs_msshf}
\end{figure*}

Fig.~\ref{fig:sst_anom_vs_msshf} shows plots for June and December derived from the lagged correlations showing only points with significant correlation at zero lag, and where the correlation magnitude peaks when SST leads the heat flux. Comparing Figs.~\ref{fig:t2merr_cov} and \ref{fig:sst_anom_vs_msshf}, we can see several qualitative similarities, and also some notable differences. We can see that both methods identify ocean-driven dynamics in the tropical eastern Pacific and tropical Atlantic in June and December, and in the Arabian Sea in June only.

However it is clear that the spatial extent of ocean-driven dynamics identified using GraphCast's errors is larger; for June, we see the spatial extent of significant correlations in Fig.~\ref{fig:t2merr_cov} (a) is noticeably greater than in Fig.~\ref{fig:sst_anom_vs_msshf} (a), particularly in the southern hemisphere. The behaviour over the northern tropical Atlantic in June also has a stronger intensity and different spatial coverage for SST-T2m in Fig.~\ref{fig:t2merr_cov} (b) compared with Fig.~\ref{fig:sst_anom_vs_msshf} (a). 
For December, larger areas of significant correlation are seen in the North Pacific and North Atlantic in Fig.~\ref{fig:t2merr_cov} (c), and in the tropical eastern Pacific, Kuroshio Extension, southern Atlantic, and off the western coast of Australia in Fig.~\ref{fig:t2merr_cov} (d).

Around the Kuroshio Extension in June, whilst the lagged correlations in Fig.~\ref{fig:sst_anom_vs_msshf} (a) identify an area of weak ocean driving concentrated nearer the coast, Fig.~\ref{fig:t2merr_cov} (b) shows positive correlation extending away from the coastline. This positive correlation (which mean a positive SST anomaly has a cooling effect on the 2-metre temperature) could indicate an area where there is a nonlinear feedback mechanism, which the lagged correlations may not be able to identify. There are likewise positive correlations in the Bay of Bengal in June [Fig.~\ref{fig:t2merr_cov} (b)], and around the Agulhas Return Current and Gulf of Mexico in December [Fig.~\ref{fig:t2merr_cov} (d)]. In contrast, the lagged correlation analysis only identifies low correlations in these areas.

There are also interesting areas of positive correlation seen for the lagged correlation results, in the eastern tropical Pacific in Fig.~\ref{fig:sst_anom_vs_msshf} (a), and over the Pacific, Atlantic and Southern Ocean in Fig.~\ref{fig:sst_anom_vs_msshf} (b). These indicate where high SST anomalies are associated with more heat flux going into the ocean. These appear to be regions for which the atmosphere is driving the dynamics; the maximum correlation for these points peaks when the SSTs lag the surface heat flux (not shown), which is what would be predicted for atmosphere-driven dynamics with a lagged response \citep{storch_signatures_2000}.

There also appear to be differences in importance for the north Pacific and Labrador Sea in June, with Fig.~\ref{fig:t2merr_cov} (a) and (b) assigning a relatively higher importance to these two eddy-active regions than Fig.~\ref{fig:sst_anom_vs_msshf} (a). This underlines the key role of nonlinear eddy-mediated coupling in these regions, that departs from large-scale linear dynamics. This may be also in part due to the presence of sea ice which is not accounted for by GraphCast.
In December, Fig.~\ref{fig:t2merr_cov} (d) identifies the southern tropical Atlantic as important for errors in T2m, with little information taken from the equatorial Atlantic, whilst Fig.~\ref{fig:sst_anom_vs_msshf} (b) shows the opposite, with correlations only present in the equatorial Atlantic.

Overall, we see that using lagged correlation analysis produces a very different picture of the locations and characteristics of ocean-driven dynamics. The regions of the ocean surface that are correlated with GraphCast's errors have a much larger spatial extent, and assign more importance to areas such as the northern Pacific and Labrador sea in June, and the southern tropical Atlantic in December. There are also areas of significant positive correlation between seen over areas such as the Kuroshio Extension in June and Agulhas Return Current in December; these positive correlations, where GraphCast has a warm bias for positive SST anomalies, may indicate a nonlinear feedback mechanism such as cloud formation that acts to cool surface temperature.

\subsection*{Dependence on climatology and interannual variability}

\begin{figure*}
[!ht]
\centering
    \includegraphics[width=17.8cm]{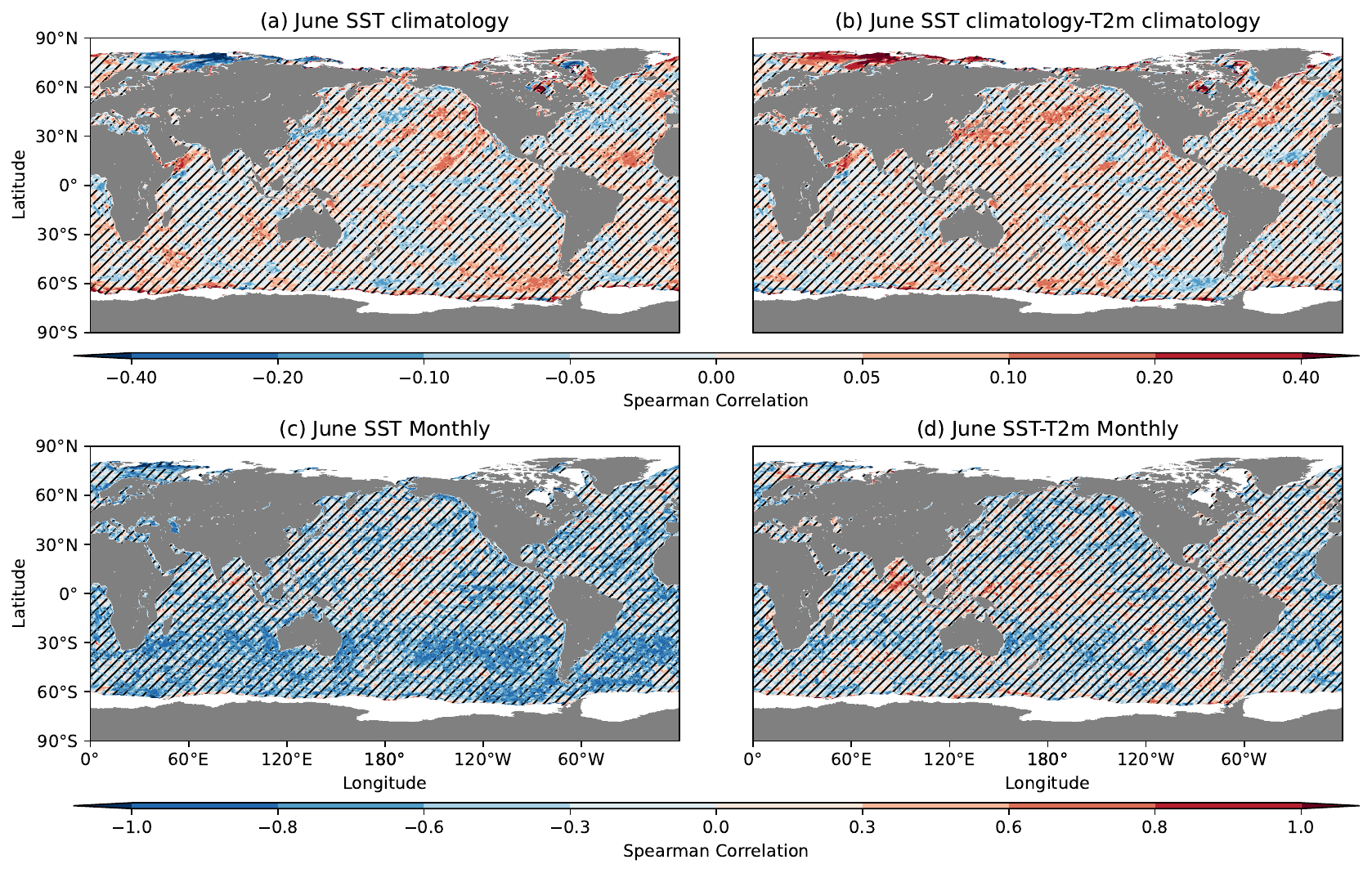}
       \caption{Top row: Spearman correlation between GraphCast's errors and sea surface variables. Hatching indicates where the correlation is not significant according to a two-tailed t-test at the 95\% confidence level. (a) SST climatology for June (b) SST climatology minus T2m climatology for June. Bottom row: Spearman correlation between GraphCast's errors aggregated to a monthly level and (c) monthly SST and (d) monthly SST-T2m, for June. }
    \label{fig:climatology_correlations}
\end{figure*}

An important point to consider is whether any of the correlations seen in Fig.~\ref{fig:t2merr_cov} are due to climatological variations of the SST. Since GraphCast receives information about which day of the year it is (through both an explicit ``day of year" variable and top-of-atmosphere radiation), we hypothesise that it is able to implicitly learn a 6-hourly climatology of the ocean, based on the average observed changes in 2m-temperature for different times of year. To test this hypothesis, we calculate correlations of SST and SST-T2m with $\varepsilon_T$ when SST and SST-T2m are replaced with the mean hourly climatology, and assess the difference in correlation compared to Fig.~\ref{fig:t2merr_cov}. If the error covaries with the SST and SST-T2m \emph{climatology}, then this suggests that GraphCast has no knowledge of the climatology. However an absence of correlations suggests that GraphCast has learnt this climatological information in some way. Representing the hourly climatology of the SST and T2m as $\text{SST}_{\text{clim}}, \text{T2m}_{\text{clim}}$, we show the correlations of GraphCast's errors with $\text{SST}_{\text{clim}}$ and $\text{SST}_{\text{clim}}-\text{T2m}_{\text{clim}}$ in Fig.~\ref{fig:climatology_correlations} (a) and (b), for June months only. Comparing the correlations of $\text{SST}_{\text{clim}}$ [Fig.~\ref{fig:climatology_correlations} (a)] with Fig.~\ref{fig:t2merr_cov} (a), we see that there are substantially fewer areas of significant correlation, suggesting that the climatology provides no new signal to GraphCast, and so in some sense GraphCast has built an internal representation of the sea surface temperature climatology. There are some remaining patches of significant correlation, particularly in the southern Pacific, Atlantic and Indian Ocean. This suggests that, for these areas, the model has not learned an accurate climatological representation, or has learned a different one than that used in this study. The correlation for $\text{SST}_{\text{clim}}-\text{T2m}_{\text{clim}}$ in Fig.~\ref{fig:climatology_correlations} (b) also shows a substantial reduction in the areas of significant correlation compared to Fig.~\ref{fig:t2merr_cov} (b), suggesting that GraphCast has learned some of the climatological heat fluxes from the ocean. Similar results are seen for the December months (not shown).

Similarly it is interesting to separate the effects of interannual and intraannual variability, to assess the timescale of the ocean dynamics that are driving the atmosphere. Some indication of this is given by the plots in Fig.~\ref{fig:climatology_correlations} (c) and (d) for June months only, where the data is first aggregated on a monthly level before correlations are calculated. Whilst this means calculating correlations over only 10 data points per grid cell, it serves as a useful indicator of whether the correlations seen in Fig.~\ref{fig:t2merr_cov} are purely driven more by interannual variability or intraannual variability. The results show that there are some areas in the southern Pacific which seem to show significant correlation between SST and $\varepsilon_T$, however there are very few areas with high significance for SST-T2m. This therefore indicates that most of the behaviour we are seeing is driven by the intraannual variability of the ocean. Similar results are seen for December (not shown).

\subsection*{Predicting errors using a simple model}
\label{sec:linear_model}

To shed light on the influence of SST on GraphCast's errors, we fit a linear model to predict these errors using only variables that are relevant to the air-sea interaction. We then test whether a significant improvement in GraphCast's errors can be obtained by removing the predicted errors from the forecast. The purposes for this are threefold; firstly it allows us to see how multiple variables interact, allowing us to see their relative importance in different regions. Secondly, it allows an estimation of the magnitude of the errors due to missing sea surface information, relative to the other errors in GraphCast. Thirdly, it is possible that we may improve the skill of GraphCast.

We use the June data between 2004-2011 as training data, June 2012 for validation, and June 2013 for testing. For hyperparameter tuning, data is regridded to 4\textsuperscript{o}, and a linear regression model fitted for each grid cell individually. All variables are normalised before fitting, in order to ensure that the relative sizes of the linear slope coefficients give the relative importance of each variable. We use the area-weighted root mean square error (AWRMSE), to match the loss function used in GraphCast's training, and only consider sea points between 50\textsuperscript{o}S-65\textsuperscript{o}N in order to exclude polar areas influenced by sea ice. We tried several linear models, including ridge regression, and Bayesian regression, finding that Lasso regression with a regularisation parameter of 0.001 performed the best. Using this optimal parameter, we then fitted a model at 1\textsuperscript{o} resolution.

\begin{figure*}[!ht]
    \centering
    \includegraphics[width=17.8cm]{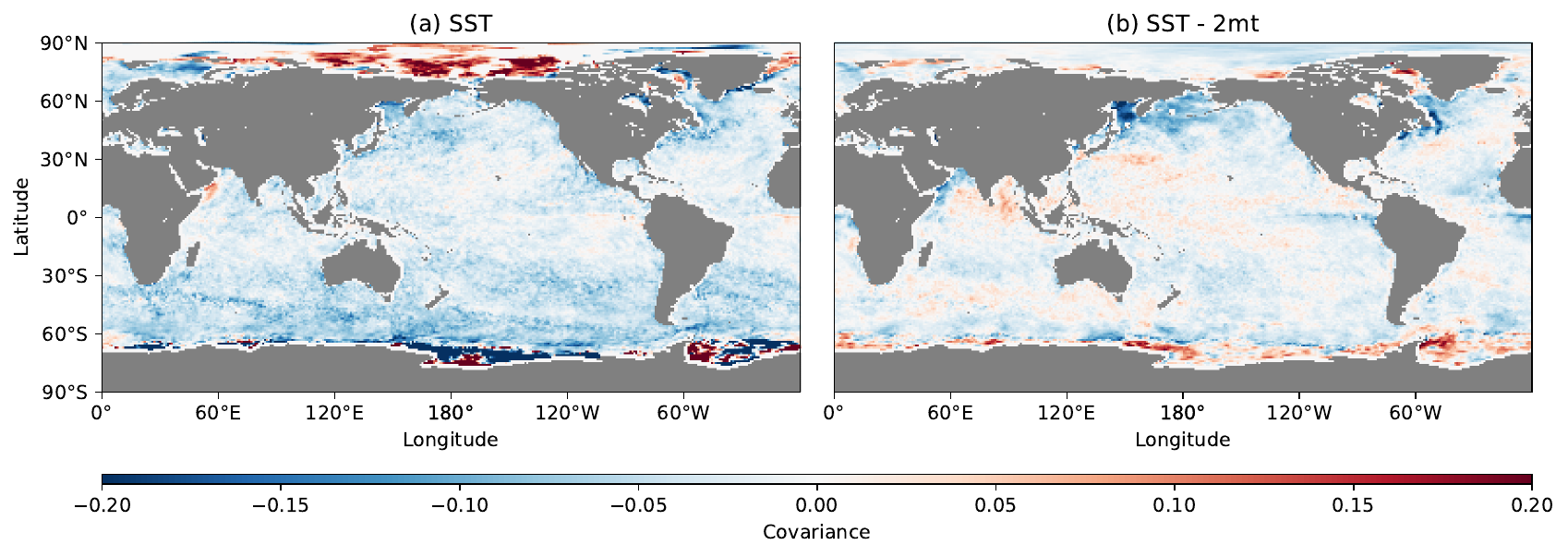}
    \caption{Linear model coefficients (using a lasso model with $\alpha=0.0001$) for June for (a) the SST input field and (b) the SST-T2m input field. In this case the model is fitted with all variables simultaneously, so where the model coefficients are zero indicates a point where that variable is not important for the prediction.}
    \label{fig:lasso_daily}
\end{figure*}

The resulting coefficients for the optimal 2-metre temperature error are shown in Fig.~\ref{fig:lasso_daily} for June. Here we can see similar relationships as shown in the correlation plots in Figs.~\ref{fig:t2merr_cov} (a) and (b), particularly in the northern Pacific, eastern Central Pacific, and Gulf Stream areas. SST-T2 is particularly important in the Gulf Stream, Bering Sea, and in the tropical Pacific and Atlantic.

In order to evaluate the efficacy of this predictive model, we evaluate its performance on unseen test data (June 2013). Since the linear model learns both an intercept (corresponding to an overall average bias observed for the particular month) and gradients for each predictive variable, we must make sure this comparison takes into account the effects of learning this intercept. Therefore, we also train a Lasso model with regularisation parameter of 100, which just learns the intercept, as a baseline for comparison.

On the test set this results in AWRMSE estimates of $0.333K$ for the baseline model and $0.331K$ on the optimal model. Therefore we estimate the SST contribution to the error over sea points to be around $1\%$ of the total. Using bootstrapping with 1000 samples we estimate the p-value of this result as $p=0.03$, suggesting the 1\% decrease in error is statistically significant.

\begin{figure*}[!ht]
    \centering
    \includegraphics[width=17.8cm]{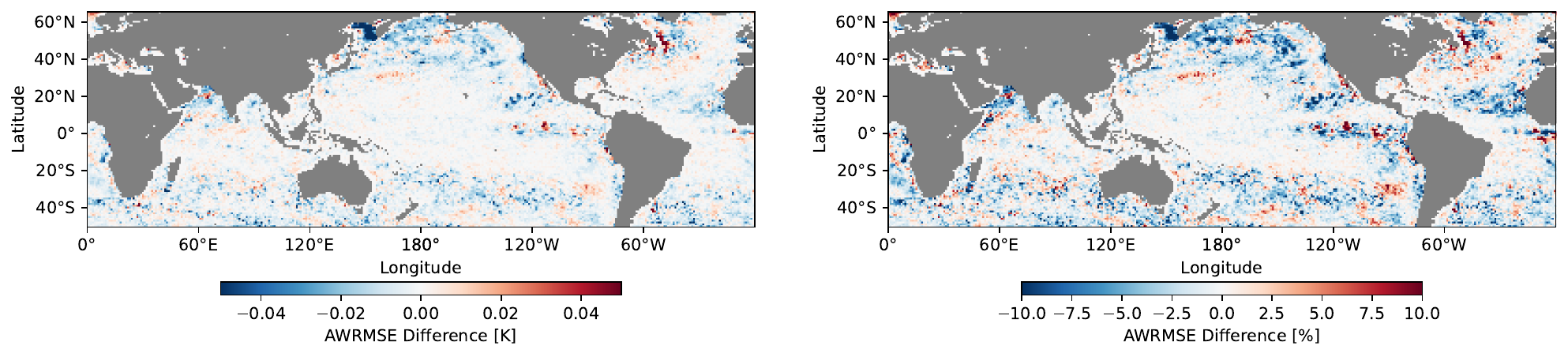}
    \caption{Difference in AWRMSE by grid cell between the baseline model with $\alpha=100$ and the optimal model with $\alpha=0.001$ (a) AWRMSE difference and (b) percentage AWRMSE difference for each grid cell. Negative (blue) values indicate where the error of the optimal model is lower (i.e. where the optimal linear model has good predictive performance).}
    \label{fig:lasso_awrmse}
\end{figure*}

The performance of the model by grid cell is also shown in Fig.~\ref{fig:lasso_awrmse}, as aboslute differences (panel a) and percentage differences (panel b). From panel (a) the largest abolsute improvements are in the northern Pacific, particularly in the Sea of Okhotsk, where the AWRMSE reduction reaches around 0.1K. From panel (b) we see that the largest percentage improvements are in the northen and eastern Pacific, and tropical Atlantic. There are also areas where the model increases the error, such as in the Labrador Sea, and in some areas of the eastern tropical Pacific; this could be due to the relatively small time period used for training and validation, such that the model has overfitted on particular features or positions of ocean currents.

\section*{Discussion}

In this work we have presented a novel application of machine-learnt weather prediction (MLWP) models; by analysing the errors of one such model (GraphCast), we can identify areas where the air-sea interaction is driven by ocean variability at short timescales. This has the potential to offer different insights compared to conventional approaches, whereby either a simplified linear model is assumed, or a general circulation model (GCM) is run with static sea surface variables. Compared to the linear model approach, our approach relaxes the assumption that the dynamics is linear, allowing a richer set of interactions to be captured. Compared to the GCM approach, our approach avoids having to accurately model the complex turbulent fluxes between the atmosphere and ocean, and instead leverages observed atmosphere-ocean variability. A limitation of this approach are that we are confined to the timescales that can be accurately forecast by GraphCast (up to around 14 days).

We demonstrate this approach by exploring the Spearman correlation between GraphCast's errors at 24h lead time, and variables relevant to the air-sea interaction. This reveals a seasonally changing dependence on sea surface temperatures, consistent with a lagged oceanic response to warming over the summer, as well as high correlations in areas associated with oceanic vertical eddy heat transport. A strong relationship is identified between GraphCast's errors and SST-T2m in the northern Pacific, Arabian Sea, northern tropical Atlantic, and eastern tropical Pacific in June. In December, the relationship is strongest in the eastern tropical Pacific, southern tropical Atlantic, and around the Kuroshio Extension. Correlations that indicate a cooling of T2m with positive SST anomalies are also seen, most notably in June around the Kuroshio Extension, and in December around the Agulhas Return Current; these may indicate nonlinear feedback mechanisms, such as cloud formation, that produce a net cooling effect with positive SST anomalies. 

A comparison with the typical lagged correlation approach reveals that, whilst there are some similarities, the regions of ocean-driven dynamics indicated by the lagged correlations have a much reduced spatial extent, and assign less importance to areas such as the northern Pacific and Labrador sea in June, and the southern tropical Atlantic in December. Additionally, the lagged correlations do not indicate the same behaviour identified over the Agulhas Return Current and Kuroshio Extension in their respecitve summers. This indicates that our novel method could be identifying interactions that violate the assumptions of the simple linear models underpinning the lagged correlations.

By calculating the same covariances using climatological ocean variables, we gain insight into the representation of the ocean that GraphCast learns. The substantial reduction in significant correlations with climatological data indicates that GraphCast has in some sense built a representation of the ocean climatology. We stress that this must be a nonlinear baseline climatological coupling, as it is derived from observations of the true atmospheric evolution including its feedbacks. Studying the average behaviour of GraphCast could therefore reveal this baseline climatology --- a topic of future study. Similarly, we explore whether these errors in GraphCast are due to the interannual or intraannual variability of the ocean, through calculating covariances with monthly data; this analysis reveals that most of the covariance identified through this technique is driven by intraannual variability, i.e., daily timescale ocean variability. 

Having learnt the relationship between GraphCast's errors and the sea surface variables, we then investigate how this relationship can be used to improve GraphCast's errors at short timescales. To do so, we build a simple linear model that predicts the errors given SST and SST-T2m. This results in an overall 1\% reduction in area-weighted RMSE (AWRMSE) at 6 hour lead times, significant at the 95\% confidence level. To our knowledge this represents the first quantitative estimate of the role of the ocean on atmospheric variability. Whilst it is a small effect, consistent with the timescales and magnitudes of variability in the atmosphere, we expect this effect to grow with lead time as the effects of the ocean become more important for forecast skill. Particularly high improvements in error are seen in the northern Pacific, with reductions in AWRMSE of up to $\sim0.1\text{K}$ seen in the Sea of Okhotsk. We also note that an improvement in forecast skill of a few percent is what could typically be expected after extensive parametrisation development \citep{haiden_evaluation_2014}.

There are several promising extensions of this approach; it would be interesting to explore how the errors associated with missing sea surface information change as GraphCast's forecast lead time is increased; since we would expect these errors to grow, there may also be potential to improve GraphCast's forecast error by a reasonable amount. Given some evidence of GraphCast under-predicting storm intensity \citep{charlton-perez_ai_2024}, and demonstrations of improvements of storm intensity for a NWP model coupled to an ocean \citep{mogensen_tropical_2017, vellinga_evaluating_2020}, it would be interesting to explore patterns of SST correlations with storm intensity error. It would also be interesting use this technique to explore interactions between the atmosphere and other slow moving earth system components, such as the land and cryosphere. Given that several other MLWP models, such as Pangu-Weather \citep{bi_accurate_2023} and FourCastNet \citep{kurth_fourcastnet_2023}, predict 2-metre temperature without taking sea surface temperature as input, it would be very interesting to explore the similarities and differences in how these model errors relate to sea surface temperature. Recent work has also demonstrated features in GraphCast's latent dimensions that relate to sea ice extent and ocean activity \citep{macmillan2025towards}; this technique could complement our approach to reveal further insights about how GraphCast internally represents the sea surface.

In summary, we have presented a way to use machine-learnt atmospheric emulators to gain insight into the nature of the air-sea interaction at short timescales. This approach can compliment existing techniques in helping us identify and understand the complex interactions and feedback mechanisms at the air-sea boundary at short timescales. 

\section*{Methods}

\subsection*{Data}
\label{sec:data}

GraphCast produces forecasts at 6 hour intervals, at 0.25\textsuperscript{o} resolution, using ERA5 variables as inputs. GraphCast data is generated over the period 2004-2013, using forecasts initialised every 6 hours and rolled out to a lead time of 24hrs, in order to remove diurnal effects. Note that, since this is not an assessment of GraphCast's skill, but instead an investigation into how GraphCast's errors correlate with the missing ocean information, it is not necessary to select a time period that lies outside GraphCast's training period.

ERA5 data \citep{hersbach_era5_2020} is used as ground truth, since this is the data that GraphCast is trained with. To reduce the data volume, and because GraphCast uses a 6hr time step, ERA5 data is sampled at 6 hour intervals. Since ERA5's SST field is set to a constant under sea ice, we mask out any areas for which the sea ice cover is greater than 0.15; the significance tests are then calculated using the total number of unmasked data points as the number of degrees of freedom.

GraphCast errors are taken as instantaneous values at the forecast lead time, whilst the ERA5 data is aggregated to a daily level, including a lag of 6 hours. This lag is to ensure that we do not include the true 2-metre temperature in the inputs. For example, if the forecast target time is 1990-01-02 12:00, then the ERA5 data is the average of the data at 1990-01-01 12:00, 1990-01-01 18:00, 1990-01-02 00:00, and 1990-01-02 06:00. 

To calculate the lagged correlations in Fig.~\ref{fig:sst_anom_vs_msshf}, we use ERA5 data aggregated to a daily resolution. The mean surface heat flux is defined as the sum of the sensible and latent heat fluxes, as used in e.g.~\cite{bishop_Scale_2017}. The climatological data used is the hourly climatology of ERA5, calculated over 1990-2019, taken from the WeatherBench datasets \citep{rasp_weatherbench_2024}.

\subsection*{Atmosphere-ocean interaction}

\label{sec:atm_ocean_interaction}
Turbulent heat fluxes between atmosphere and ocean are typically characterised using the bulk formulae \citep{stensrud_wateratmosphere_2007}:
\begin{align}
Q_H &= \rho_a C_H c_a (T_o - T_s) |\mathbf{u}_z |\\
Q_L &=  - L_v  \rho_a C_E (q_s - q_z) |\mathbf{u}_z| = - L_v E
\end{align}
Where $\rho_a$ is the air density, $c_a$ the specific heat capacity of moist air, $L_v$ the latent heat of vaporisation, $\mathbf{u}_z$ the horizontal wind velocity at height $z$ (typically taken to be relative to the surface currents). $T_o$ is the sea surface temperature (either bulk or skin temperature), whilst $T_s$ is the temperature of air near the ocean-air interface (which is in general not equal to the bulk air temperature $T_a$). $C_D, C_E$ are the bulk coefficients for sensible heat and evaporation, estimated using Monin–Obukhov similarity theory \citep{stensrud_wateratmosphere_2007}. $q_s$ is the saturation specific humidity of air at temperature $T_o$ and pressure equal to the sea level pressure. In addition to the turbulent heat fluxes, there are also long-wave radiation fluxes between the ocean and atmosphere.

Following the approach in \cite{schopf_modeling_1985} we treat the atmosphere as a single well-mixed layer of height $H_a$ with bulk temperature $T_a$ and surface temperature $T_{a,0}$. The ocean is similarly treated as a single well-mixed layer of depth $H_o$ with bulk temperature $T_o$ and surface temperature $T_{o,0}$. Feedback effects on surface winds due to the SST and cloud radiative feedback are ignored.
Combining the above equations, and accounting for incoming shortwave radiation and dynamical effects through terms $R_a, R_o$ and $F_a, F_o$ respectively, gives the following coupled equations for the evolution of atmospheric and oceanic temperature near the surface (ignoring the effects of evaporated water vapour on the atmosphere, as the effect is small compared to the sensible heat flux):
\begin{align}
    \rho_a H_a c_a \frac{dT_a}{dt} &= \rho_a H_a C_H c_a (T_{o,0} - T_{a,0}) |\mathbf{u}_z | + \varepsilon_o \sigma_B T_{o}^4 \nonumber\\
    &- 2\varepsilon_a \sigma_B T_{a}^4 + R_a + F_a \label{eq:cpl_atm}\\
    \rho_o H_o c_o \frac{dT_o}{dt} &= - \rho_a H_a C_H c_a (T_{o,0} - T_{a,0}) |\mathbf{u}_z | - \varepsilon_o\sigma_B T_{o}^4 \nonumber\\
    &+ \varepsilon_a \sigma_B T_{a}^4  - L_v  \rho_a C_E (q_s - q_z) |\mathbf{u}_z| \nonumber \\
    &+  R_o + F_o \label{eq:cpl_ocn}
\end{align}

From this we can see that the terms affecting $T_a$ that depend on the ocean state are predominantly functions of $(T_{o,0} - T_{a,0})$ and $T_o^4$. If we assume that the surface and bulk temperatures are linearly related, then this leads to a function of $(T_{o} - T_{a})$, $T_o$ and $T_o^4$. 

Many studies of the atmosphere-ocean interaction, such as \cite{wu_local_2006} and \cite{bishop_Scale_2017}, use a linearised version of the above equation, shown in \cite{barsugli_basic_1998}. To arrive at this, an assumption is first made that the combined latent heat flux and sensible heat flux terms can be written as linear functions of $T_o, T_s$. Then each temperature term is expanded as $T_x = \bar{T}_x + T_x'$, where $ \bar{T}_x$ is the mean state and $T_x'$ is the anomaly, assumed to be small compared to the mean state. It is further assumed that $R_o'=R_a'=0$, and the dynamical terms $F_a', F_o'$ are modelled as white noise processes. Under these assumptions, the Eqs.~\eqref{eq:cpl_atm} and \eqref{eq:cpl_ocn} become:
\begin{align}
   \frac{dT'_a}{dt} &\simeq  \lambda_{sa}(T_o' - T_a') + \lambda_a T_a'+ F_a' \\
   \frac{dT_o}{dt} & \simeq  
    \lambda_{so}(T_a' - T_o') +  \lambda_o T_o'  + F_o' 
\end{align}
where $\lambda_{sa}$ and $\lambda_{so}$ parameterise the total effect of the linearised sensible and latent heating (assumed to be constant in time). $\lambda_a, \lambda_o$ parameterise the net effect of long-wave radiative heating and cooling, as well as the difference between $T_a$ and $T_{a,0}$, and the difference between $T_o$ and $T_{o, 0}$ (where it is assumed they follow a linear relationship).

\subsection*{Significance testing}
\label{sec:sig}
We test for correlations being significantly high or low by using a two-sided t-test of the test statistic $r_{a} ( n-2)^{1/2} / (1-r_{a}^2)^{1/2}$ \citep{storch_statistical_2002}, where $n$ is the number of data points used to calculate the result. To calculate the significance of the error reduction from constructing a linear model to predict errors based on sea surface temperature, we use bootstrapping in order to calculate a mean and standard error of the AWRMSE for GraphCast and GraphCast corrected by a linear model. The significance is then calculated using a two-sample t test \citep{storch_statistical_2002}.

\section*{Data Availability}

The processed ERA5 and GraphCast error datasets analysed during the current study are available in the Zenodo repository, \url{https://doi.org/10.5281/zenodo.18470825}. The full GraphCast error dataset generated during the current study is not publicly available due to storage limitations, but is available from the corresponding author on reasonable request. The ERA5 datasets used are available in the Copernicus Climate Change Service, Climate Data Store, (2023), for registered users. The dataset used is the ERA5 hourly data on single levels from 1940 to present (DOI: 10.24381/cds.adbb2d47). Accessed July-August 2025. The GraphCast model code and weights are available via \url{https://github.com/google-deepmind/graphcast}. Climatology data is taken from the WeatherBench 2 dataset \citep{rasp_weatherbench_2024}, more details available at \url{https://weatherbench2.readthedocs.io/}.
\section*{Code Availability}

The underlying code for this study is available at 
\url{https://github.com/bobbyantonio/graphcast_ocean_interaction}.

\section*{Acknowledgements}
This publication is part of the EERIE project funded by the European Union (Grant Agreement No 101081383). Views and opinions expressed are however those of the author(s) only and do not necessarily reflect those of the  European Union or the European Climate Infrastructure and Environment Executive Agency (CINEA). Neither the European Union nor the granting authority can be held responsible for them. This work was funded by UK Research and Innovation (UKRI) under the UK government’s Horizon Europe funding guarantee (grant number 10049639).

\section*{Author Contributions}
All authors contributed to the study conception and design. BA generated the datasets and performed the analysis. The first draft of the manuscript was written by BA and all authors edited subsequent versions of the manuscript.

\section*{Competing Interests}

All authors declare no financial or non-financial competing interests.

\bibliographystyle{apalike}
\bibliography{references}

\end{document}